\documentstyle[12pt,aaspp4]{article} 
\def\tempest%
{\begin{array}{ccc} 
1 & 1 & 1 \\ 
1 & 1 & 1 \\ 
4 & 3 & 8 
\end{array}}

\def\kms{{\rm km}\,{\rm s}^{-1}} 
 
\def\rel{{\rm rel}}

\def\e{{\rm E}}

\def\bpi{\hbox{$\pi\hskip-6.6pt\pi$}}
\def\bmu{\hbox{$\mu\hskip-7.5pt\mu$}}

\begin{document}

\title{A Natural Formalism for Microlensing} 
\author 
{Andrew Gould}
\affil{Ohio State University, Department of Astronomy, 
140 W.\ 18th Ave., Columbus, OH 43210, USA; gould@astronomy.ohio-state.edu} 

\begin{abstract} 

	If the standard microlensing geometry is inverted so that the
Einstein ring is projected onto the observer plane rather than the source
plane, then the relations between the observables $(\theta_\e,\tilde r_\e)$
and the underlying physical quantities $(M,\pi_\rel)$ become immediately
obvious.  Here $\theta_\e$ and $\tilde r_\e$ are the angular and projected
Einstein radii, $M$ is the mass of the lens, and $\pi_\rel$ is the lens-source
relative parallax.  I recast the basic formalism of microlensing in light
of this more natural geometry and in terms of observables. I then find that the
relations between observable and physical quantities assume an 
exceptionally simple form.  In an appendix, I propose a set of notational 
conventions.


\end{abstract} 

\section{Introduction} 

	The geometry of point-lens microlensing (Einstein 1936; Refsdal 1964; 
Paczy\'nski 1986) is so simple that students can derive all the basic
results in a few hours.  Nonetheless, this geometry has never been boiled
down to its essence: the relationship between the underlying physical
quantities and the observables.  In particular, the ``Einstein ring radius''
$r_\e$, a central concept in the usual formulation, is not directly observable
and has not been observationally determined for even one of the $\sim 500$
microlensing events observed to date.  There appear to be three reasons that
the natural geometric formulation has not been developed.  First, the 
standard geometry is already so trivial that further simplification has
not seemed worthwhile.  Second, the theory of microlensing was already quite
developed before it was realized what the observables were, and until very
recently, the prospects were poor for measuring these observables except in a
handful of events.  Third, the original impulse to microlensing searches
was to probe the dark matter. This focused attention on the optical depth
(a statistical statement about the ensemble of events) and secondarily
on the Einstein timescale $t_\e$, which of the three observables is the one
that has the most convoluted relation to the underlying physical parameters.

	However, with the prospect astrometric microlensing it is now
possible that a second observable, the angular Einstein radius $\theta_\e$,
will be routinely measured (Boden, Shao, \& Van Buren 1998; Paczy\'nski 1998).
Moreover,  if these astrometric measurements are carried out by the
Space Interferometry Mission (SIM) in solar orbit, then comparison of
photometry from SIM and the ground will yield a third observable, the
projected Einstein radius $\tilde r_\e$ (Refsdal 1966; Gould 1995; Gould \&
Salim 1999).  Hence, it is now appropriate
to reformulate the microlensing problem in terms of these observables.

\section{Geometry}

	The upper panel of
Figure \ref{fig:one} shows the standard presentation of microlensing
geometry (e.g.\ Fig.\ 3 from Gould 1996).  The observer (O), lens (L) of
mass $M$, and source (S) are aligned.  The 
light is deflected by an angle $\alpha$ given by the Einstein (1936) formula
\begin{equation}
\alpha = {4 G M\over r_\e c^2},
\label{eqn:alphadef}
\end{equation}
where $r_\e$ is the Einstein radius.  It arrives at the observer displaced
by an angle $\theta_\e$ from the true position of the source.  In this case,
the source is therefore imaged into a ring.  The size of this ring projected
onto the source plane is $\hat r_\e$.  More generally, the 
alignment will not be perfect, and the axial symmetry will be broken.  Hence, 
there will be two images rather than a ring.  However, even in this more 
general case the
Einstein ring provides a natural scale to the problem.  

	The lower panel of Figure \ref{fig:one} basically inverts the geometry 
of the upper panel and thereby focuses attention on the observer rather
than the source.  This seems like a trivial change but it has two
advantages.  First, the quantities shown at the right, $\theta_\e$ and
$\tilde r_\e$ are the observables.  To date, $\theta_\e$ has been measured
for only 4 events (Alcock et al.\ 1997; Albrow et al.\ 1999,2000; 
Afonso et al.\ 2000), all by using the source as an ``angular ruler''
(Gould 1994a; Nemiroff \& Wickramasinghe 1994; Witt \& Mao 1994).  
Similarly, $\tilde r_\e$ has been determined
for only about a half dozen events (Alcock et al.\ 1995; Bennett et al.\ 1997; 
Mao 1999).  For all of these, $\tilde r_\e$ was found by measuring the
deviation of the light curve induced by the Earth's motion (Gould 1992).
The amplitude of this deviation is proportional to 
$\pi_\e\equiv {\rm AU}/\tilde r_\e$.  The measurements of both $\theta_\e$
and $\tilde r_\e$ have required
special conditions (a caustic crossing for $\theta_\e$ and an event lasting
a large fraction of a year for $\tilde r_\e$), which is why so few of these
``observables'' have actually been observed.  However, as mentioned above,
both $\theta_\e$ and $\tilde r_\e$ could be measured routinely in the future.

	The second reason for inverting the standard geometry is that doing so
makes transparent the relation between the observables and the underlying
physical variables: the product of $\theta_\e$ and $\tilde r_\e$ is
essentially the Schwarzschild radius of the lens, and their ratio is
essentially the lens-source relative parallax.  Using the small angle 
approximation, one sees immediately from the lower panel of 
Figure \ref{fig:one} that
$\alpha/\tilde r_\e = \theta_\e/r_\e$, or
\begin{equation}
\theta_\e \tilde r_\e = \alpha r_\e = {4G M\over c^2}.
\label{eqn:timeseq}
\end{equation}
Next, from the exterior-angle theorem
\begin{equation}
\theta_\e = \alpha - \psi = {\tilde r_\e\over d_l} - {\tilde r_\e\over d_s}
= {\tilde r_\e\over d_\rel},
\label{eqn:diveq}
\end{equation}
where $d_l$ and $d_s$ are the distances to the lens and source, and
$d_\rel^{-1}\equiv d_l^{-1}-d_s^{-1}$.  Note that equation (\ref{eqn:diveq})
can be written more suggestively as
\begin{equation}
\pi_\e \theta_\e = \pi_\rel,\qquad \pi_\e\equiv {{\rm AU}\over\tilde r_\e},
\label{eqn:relpi}
\end{equation}
where $\pi_\rel={\rm AU}/d_\rel$ is the lens-source relative parallax.

	Just as in astrometric parallax determinations where $\pi$ is a more 
natural way to represent the measured quantity than its inverse (distance), 
so in microlensing ``parallax'' determinations, $\pi_\e$ is more natural than
its inverse $(\tilde r_\e)$.  The reason is the same: the observable effect is
inversely proportional $\tilde r_\e$ but directly proportional to $\pi_\e$,
so the measurement errors, when expressed in terms of $\pi_\e$ exhibit more 
regular behavior.  As in the case of astrometric parallax, this feature
becomes especially important for measurements that are consistent with zero
at the few $\sigma$ level.
Indeed, in contrast to astrometric parallaxes, microlensing
parallaxes are inherently two-dimensional (Gould 1995).  That is, one
measures not only the amplitude of $\tilde r_\e$ (or $\pi_\e$) but also
the direction lens-source relative motion.  Hence one can generalize
$\pi_\e$ to a two-dimensional vector $\bpi_\e$ whose direction is that of
the lens relative to the source.  The measurement errors in $\bpi_\e$
are then easily expressed as a covariance matrix.  By contrast, there is
no natural way to generalize $\tilde r_\e$: it can be made into a vector
with the same direction $\tilde {\bf r}_\e$, but when
$\bpi_\e$ is consistent with zero, such a vector is very poorly behaved.
Moreover, in some cases 
one component of $\bpi_\e$ can be very well determined while the
other is highly degenerate (Refsdal 1966; Gould 1994b,1995), 
a situation that is easily represented using $\bpi_\e$ but unwieldy using
$\tilde {\bf r}_\e$.  (Note that while no one has ever previously 
introduced $\tilde {\bf r}_\e$, I have often discussed the closely
related projected velocity,
$\tilde {\bf v}=\tilde{\bf r}_\e/t_\e$.)

	The Einstein crossing time $t_\e$ is the only observable that at 
present is routinely observed.  While I find no fault with $t_\e$, 
considerations of symmetry with the substitution 
$\tilde r_\e\rightarrow \bpi_\e$ lead me to substitute 
$t_\e\rightarrow \bmu_\e$, where
\begin{equation}
\mu_\e\equiv {1\over t_\e},
\label{eqn:muedef}
\end{equation}
and where the direction of $\bmu_\e$ is that of the lens motion relative to the
source.
With this definition, the relative lens-source proper motion is given by
$\bmu_\rel = \bmu_\e\theta_\e$.

\section{Relations Between Observables and Physical Quantities}

	From equations (\ref{eqn:timeseq})--(\ref{eqn:relpi}), one
immediately derives
\begin{equation}
\tilde r_\e = \sqrt{4 G M d_\rel\over c^2},
\qquad \pi_\e = \sqrt{\pi_\rel\over \kappa M},
\label{eqn:tilderedef}
\end{equation}
and
\begin{equation}
\theta_\e = \sqrt{4 G M \over d_\rel c^2} = \sqrt{\kappa M\pi_\rel},
\label{eqn:thetaedef}
\end{equation}
where
\begin{equation}
\kappa \equiv {4 G\over c^2\rm AU} = {4\,v_\oplus^2\over M_\odot c^2}
\simeq 8.144\,{{\rm mas}\over M_\odot},
\label{eqn:kappadef}
\end{equation}
and $v_\oplus\sim 30\,\kms$ is the speed of the Earth.  

	How well is the coefficient ($8.14\ldots$) in $\kappa$ known?
It suffers from two sources of uncertainty.  First, the factor ``4''
in equations (\ref{eqn:kappadef}) and (\ref{eqn:alphadef}) is a prediction
of General Relativity (GR).  It's accuracy (often parameterized by $\gamma$) 
has been verified experimentally by Hipparcos, but only
to 0.3\% (Froeschle, Mignard, \& Arenou 1997).  However, if GR is assumed to
be exact, then this coefficient can be determined
as accurately as $(v_\oplus/c)^2$, which should be known from pulsar timing
and solar-system radar ranging to at least nine significant digits.

	In astrometric microlensing measurements, one automatically recovers
the parallax and proper-motion of the source, $\pi_s$ and $\bmu_s$ 
(Boden et al.\ 1998; Gould \& Salim 1999).  Hence, the observables 
are $\bmu_\e$, $\bpi_\e$, $\theta_\e$, $\pi_s$ and $\bmu_s$.
When expressed in this natural form, they have a particularly simple relation
to the physical properties of the lens:
\begin{equation}
M = {\theta_\e\over \kappa\pi_\e},
\label{eqn:meq}
\end{equation}
\begin{equation}
\pi_l = \pi_\e\theta_\e + \pi_s,
\label{eqn:pil}
\end{equation}
\begin{equation}
\bmu_l = \bmu_\e\theta_\e + \bmu_s,
\label{eqn:bmul}
\end{equation}
and
\begin{equation}
{\bf v}_{\perp,l} = {\bmu_\e\theta_\e + \bmu_s\over \pi_\e\theta_\e + \pi_s},
\label{eqn:vl}
\end{equation}
where $\pi_l$, $\bmu_l$, and ${\bf v}_{\perp,l}$ are the parallax, proper 
motion, and transverse velocity of the lens.
\smallskip

{\bf Acknowledgements}: 
This work was supported by grant AST 97-27520 from the NSF.

\appendix
\renewcommand{\theequation}{\thesection\arabic{equation}}
\section{The Need for Uniform Notation}

	Microlensing suffers from a plethora of mutually 
inconsistent notational conventions.  While this poses no real problem
for veterans, it presents significant obstacles to newcomers entering the
field.  I take the opportunity of this paper (which, more than most,
concerns itself with notational issues) to try to forge a consensus.
In formulating my proposed conventions, I am influenced primarily by
prevalence of current usage, and secondarily by the need for internal
consistency.  

	I am abandoning some of my own prized notations, and I hope others
are willing to do the same in the interest of achieving a uniform system.
I will post this manuscript on astro-ph and circulate it privately to a wide
audience thereby allowing an informal ``vote'' on my proposal and corrections
to it if they seem required.  In the final published version of
the paper, I will replace this paragraph with the results of that ``vote''.

	First, all quantities associated with the size of the Einstein ring 
(in units of length, angle, time, etc.) should be subscripted with an 
upper-case roman ``E'' in conformity with ApJ conventions.  All physical 
Einstein radii should be denoted $r$.  Hence,
$\tilde r_\e$, $r_\e$, and $\hat r_\e$ for the Einstein rings in the
planes of the observer, lens, and source.  The other quanities are
$\theta_\e$ for the angular Einstein radius, $t_\e$ for the Einstein crossing
time, $\mu_\e\equiv t_\e^{-1}$, $\pi_\e\equiv {\rm AU}/\tilde r_\e$, and
the direction of $\bpi_\e$ and $\bmu_\e$ defined by the direction of the
proper motion of the lens relative to the source.

	Second, all quantities associated with position in the Einstein ring 
should be denoted by $u$, or possibly by $\bf u$ if a vector position is 
indicated.
When $\bf u$ is a vector, it must be specified whether it is the source
position relative to the lens or vice versa.  Common usage seems to conform to 
the former, and hence I adopt that.  However, keep in mind that this means that
$d{\bf u}/d t = -\bmu_\rel/\theta_\e$.

	Third, all quantities associated with the time of closest approach
to the center of the Einstein ring should be denoted by a subscript ``0''.  
Thus, $t_0$ for the time of closest approach and $u_0$ for the projected 
separation of the lens and source in units of $\theta_\e$ at time $t_0$.

	Fourth, all quantities associated with the source should be denoted
by a subscript ``$*$''.  Thus, $\theta_*$ for the angular radius of the 
source and $r_*$ for its physical radius.

	Fifth, time normalized to the Einstein crossing time should be
denoted $\tau=(t-t_0)/t_\e$.  Hence the vector position in the Einstein
ring is ${\bf u} = (\tau,u_0)$.

	Sixth, event parameters as measured from locations other than the Earth
should be subscripted, e.g., ``$t_{0,s}$'' for the time of closest
approach as seen from a satellite.  The subscript ``$\odot$'' should be 
reserved for event parameters as seen from the Sun (not in the Sun frame
but from another location).  The ``$\oplus$'' subscript should be used only 
when needed to avoid confusion.

	Finally, the reader will note that I have described different 
parameters that contain the same information, e.g., $(\tilde r_\e,\pi_\e)$
and $(t_\e,\mu_\e)$.  I expect that $\bpi_\e$ and $\bmu_\e$ will come into
use mainly in technical applications, and that the general reader of
microlesing articles will continue to find $\tilde r_\e$ and $t_\e$
to be more intuitive.  In particular, in cases where there is only
a microlensing parallax measurement, the projected velocity 
$\tilde {\bf v} = \bmu_\e/\pi_\e$ is often a substantially more
useful representation of the measurement than $\bmu_\e$ and $\pi_\e$
reported separately.  Note that in contrast to 
${\bf v}_{l,\perp} = \bmu_l/\pi_l$, which represents two components of an
intrinsically three-dimensional vector, $\tilde{\bf v}$ is intrinsically
two-dimensional and so should not be subscripted with a ``$\perp$''.

\bigskip

\newpage

\begin{figure}
\caption[junk]{\label{fig:one}
Upper panel:
standard microlensing geometry.  Bold curve shows the path of the light
from the source (S) to the observer (O) being deflected by the lens (L) 
of mass $M$.  The deflection angle is $\alpha=GM/r_\e c^2$, where $r_\e$
is the Einstein radius shown as a dashed line.  The image (I) is displaced
from the source by the angular Einstein radius $\theta_\e$ which, projected
onto the source plane, corresponds to a physical distance $\hat r_\e$.

Lower panel:
natural microlensing geometry.  Mostly the same as the upper panel
except that the Einstein radius is now projected onto the observer plane
as $\tilde r_\e$ rather than onto the source plane as $\hat r_\e$.  This
minor difference allows one to see immediately the relations between the
observables ($\theta_\e$, $\tilde r_\e$) and the physical parameters
$(M,\pi_\rel)$.  First, under the small-angle approximation, 
$\alpha/\tilde r_\e=\theta_\e/r_\e$, so 
$\tilde r_\e\theta_\e = \alpha r_\e= 4GM/c^2$.  Second, by the exterior-angle
theorem, $\theta_\e = \alpha - \psi = \tilde r_\e/d_l - \tilde r_\e/d_s$,
where $d_l$ and $d_s$ are the distances to the lens and source.  Hence,
$\theta_\e/\tilde r_\e = \pi_\rel/\rm AU$, where $\pi_\rel$ is the lens-source
relative parallax.
}
\end{figure}

\clearpage

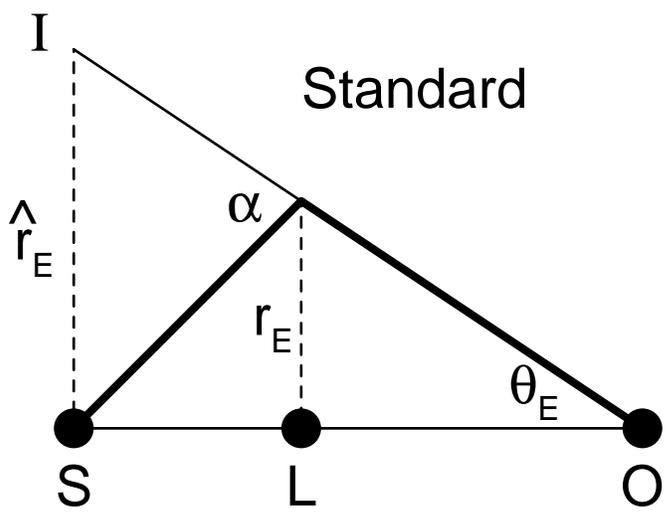
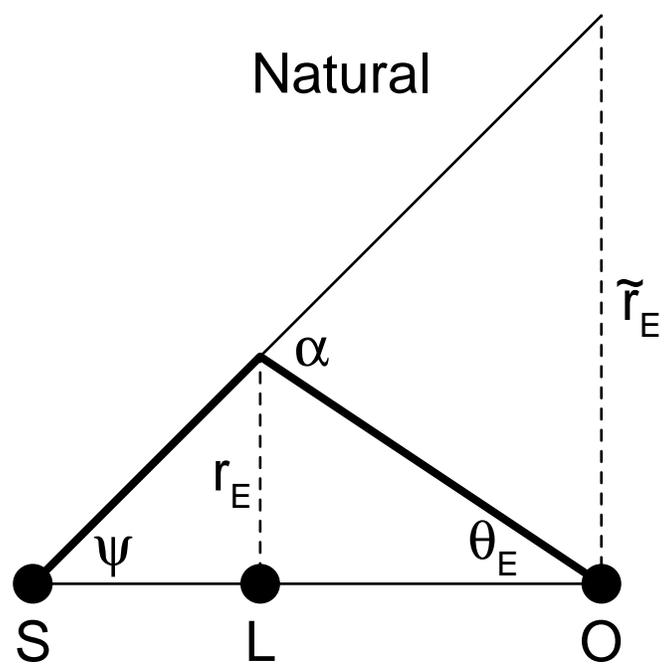
\end{document}